\documentclass[aps,prl,twocolumn,superscriptaddress]{revtex4}

\usepackage{amsmath}
\usepackage[dvips]{graphicx}
\usepackage{color}
\usepackage{hyperref}

\numberwithin{equation}{section}

\begin{document}

\title{Singlet Ground State of the Quantum Antiferromagnet Ba$_3$CuSb$_2$O$_9$}

\author{J.~A.~Quilliam}
\author{F.~Bert}
\author{E.~Kermarrec}

\affiliation{Laboratoire de Physique des Solides, Universit\'{e} Paris-Sud 11, UMR CNRS 8502, 91405 Orsay, France}

\author{C.~Payen}
\author{C.~Guillot-Deudon}
\affiliation{ Institut des Mat\'{e}riaux Jean Rouxel, UMR 6502 Universit\'{e} de Nantes, CNRS,  2 rue de la Houssini\`{e}re, BP 32229, 44322 Nantes Cedex 3, France}

\author{P.~Bonville}
\affiliation{Service de Physique de lÕ\'{E}tat Condens\'{e}, CEA-CNRS, CE-Saclay, F-91191 Gif-Sur-Yvette, France}

\author{C.~Baines}
\author{H.~Luetkens}
\affiliation{Laboratory for Muon Spin Spectroscopy, Paul Scherrer Institute, CH-5232 Villigen PSI, Switzerland}

\author{P.~Mendels}
\affiliation{Laboratoire de Physique des Solides, Universit\'{e} Paris-Sud 11, UMR CNRS 8502, 91405 Orsay, France}
\affiliation{Institut Universitaire de France, 103 bd Saint-Michel, F-75005 Paris, France}

\date{\today}

\begin{abstract}

We present local probe results on the honeycomb lattice antiferromagnet Ba$_3$CuSb$_2$O$_9$.  Muon spin relaxation measurements in zero field down to 20 mK show unequivocally that there is a total absence of spin freezing in the ground state. Sb NMR measurements allow us to track the intrinsic susceptibility of the lattice, which shows a maximum at around 55 K and drops to zero in the low-temperature limit.  The spin-lattice relaxation rate shows two characteristic energy scales, including a field-dependent crossover to exponential low-temperature behavior, implying gapped magnetic excitations. 

\end{abstract}

\keywords{}

\maketitle


In recent years, the field of geometrically frustrated magnetism has been galvanized by the discovery of several promising candidates for quantum spin liquid (QSL) physics in kagome~\cite{Mendels2007,Helton2007}, hyperkagome~\cite{Okamoto2007} and triangular lattice~\cite{Shimizu2003,Itou2008} geometries.    The most recent and surprising discoveries in this vein have come in the 6H-perovskite family, Ba$_3M$Sb$_2$O$_9$ where $M = $Cu, Ni~\cite{Zhou2011,Cheng2011}, and have generated an enormous theoretical interest~\cite{Serbyn2011,Xu2011,Rubin2012,Chen2012,Thompson2012}.  

Here we focus on the $S=1/2$ antiferromagnet Ba$_3$CuSb$_2$O$_9$, originally thought to feature a triangular lattice of Cu$^{2+}$ spins~\cite{Kohl1978} and shown by Zhou \emph{et al.} to exhibit a very surprising gapless QSL state~\cite{Zhou2011} with a lack of static magnetism down to 200 mK, well below the Weiss temperature, $\theta_W \simeq 50$ K, and a peculiar linear specific heat.  An in-depth characterization, including single crystals, of Nakatsuji \emph{et al.}~\cite{Nakatsuji2012} on the other hand has shown that the structure more likely involves triangular bilayers equally occupied by Cu and Sb that form a decorated honeycomb lattice during crystal growth as a result of electric-dipole interactions between Sb$^{5+}$-Cu$^{2+}$ `dumbbells' that map onto a frustrated Ising model.  Moreover, it has been shown that a crucial ingredient in the physics of Ba$_3$CuSb$_2$O$_9$ is the orbital degrees of freedom.  The system is Jahn-Teller (JT) active, with three equivalent local distortions and corresponding $d_{x^2-y^2}$ orbitals.  In some off-stoichiometric samples, a collective orthorhombic JT transition at $\sim 200$ K occurs, whereas stoichiometric samples maintain hexagonal symmetry, show minimal anisotropy and remain dynamic.

The lack of static magnetism in Ba$_3$CuSb$_2$O$_9$ to $T\ll \Theta_W \sim 50$ K  remains a profound result.  Neither triangular nor honeycomb nearest neighbor antiferromagnets are expected to have QSL ground states and apparent JT instabilities make it even more surprising.  Two likely proposals have emerged: either a new type of QSL on the honeycomb lattice accompanied by dynamic JT distortions or a random singlet state (RSS) driven by static JT distortions on a rather disordered lattice~\cite{Nakatsuji2012,Balents2012}.  An understanding of this system therefore hinges on a characterization of the ground state and the evolution of the susceptibility and spin fluctuations.

In this Letter, we have applied the local probe techniques, muon spin relaxation ($\mu$SR) and nuclear magnetic resonance (NMR), to Ba$_3$CuSb$_2$O$_9$. We demonstrate in this system a lack of static magnetism in zero-field,  singlet formation around 50 K and a lower temperature field-induced gap.  Our measurements help distinguish between the two possible scenarios proposed in Ref.~\cite{Nakatsuji2012} and favor a RSS.
 

The synthesis and structural characterization of the polycrystalline Ba$_3$CuSb$_2$O$_9$ samples studied in this work are described in the Supplemental Material (SM)~\cite{SM}.  Macroscopic susceptibility, $\chi_\mathrm{macro}$, measurements reveal a Weiss temperature, $\theta_W \simeq 51$ K, consistent with previous work~\cite{Zhou2011,Nakatsuji2012} and Curie constant 0.48 Kemu/mol/Oe, implying either a somewhat large $g = 2.27(1)$ (relative to ESR results, $g = 2.20$~\cite{Nakatsuji2012}) or else a 6\% surplus of Cu.  All characterization techniques employed~\cite{SM} have revealed an Sb-Cu ratio stoichiometric to well within $\pm 10\%$ indicating~\cite{Nakatsuji2012} that our samples should maintain hexagonal symmetry.

\begin{figure}
\begin{center}
\includegraphics[width=3.25in,keepaspectratio=true]{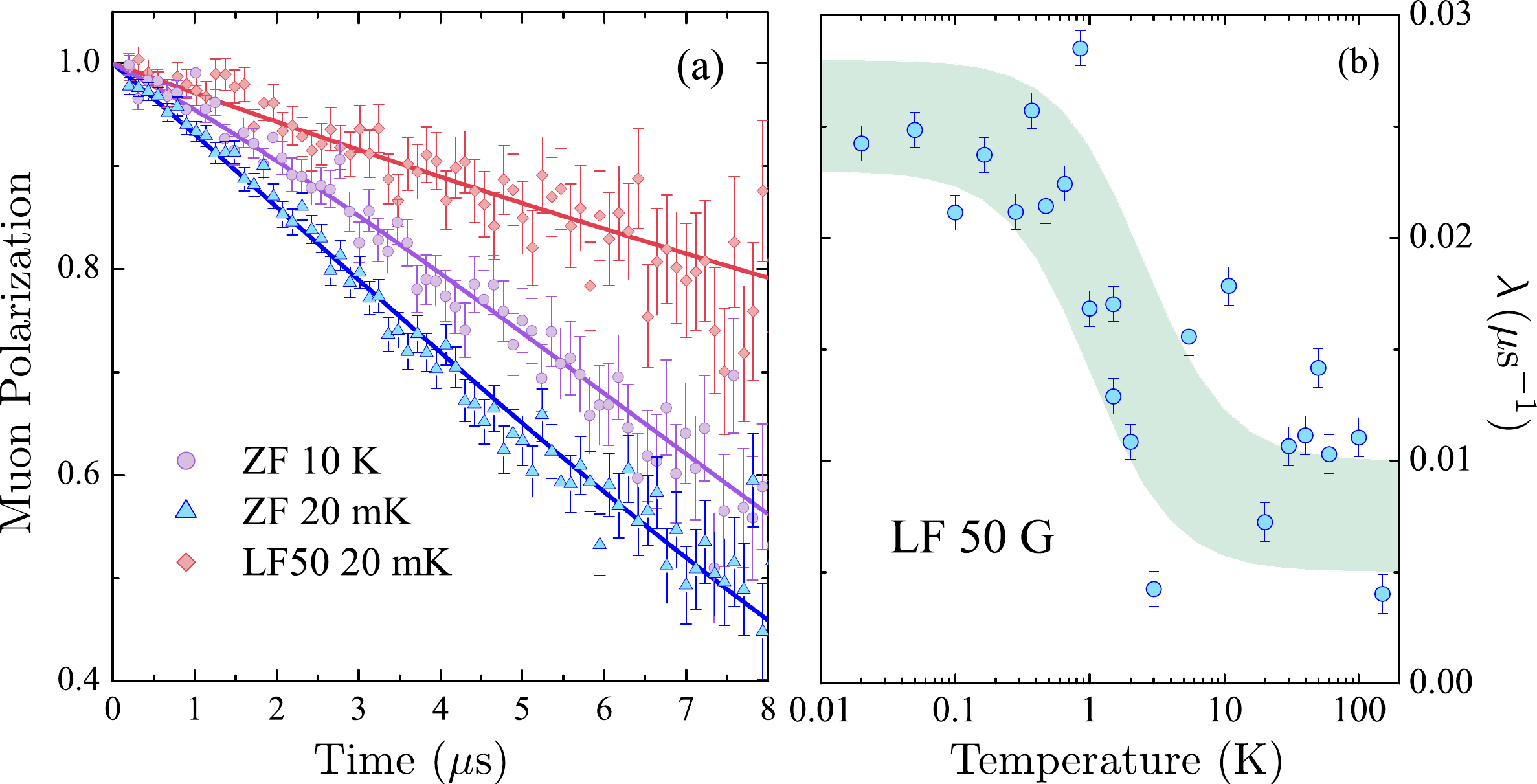}
\caption{(a) Muon polarization at $T=10$ K in zero field and at $T = 20$ mK in zero field, and longitudinal field of 50 G.  The fits are exponential relaxation, multiplied with a KT form of nuclear origin for the ZF data.  (b) Relaxation rate $\lambda(T)$ in a longitudinal field of 50 G.  The shaded region is a guide to the eye}
\label{muSR}
\end{center}
\end{figure}

To further confirm that our samples are consistent with the hexagonal phase of Ref.~\cite{Nakatsuji2012} and not the orthorhombic phase that exhibits freezing at $T_g = 100$ mK, we performed $\mu$SR experiments at PSI from 150 K down to 20 mK.  The zero field (ZF) polarization of the bulk~\cite{parimp} can be fit at all temperatures with a static Kubo-Toyabe (KT) function with $\Delta H = 0.76 \pm 0.2$ G, resulting from nuclear moments, and weak exponential relaxation of dynamical origin (see Fig.~\ref{muSR}).  The results show that there is a complete lack of spin freezing~\cite{upperbound} in Ba$_3$CuSb$_2$O$_9$ down to 20 mK in ZF, in agreement with Ref.~\cite{Nakatsuji2012}.  Applying a 50 G longitudinal field (LF) decouples the muons from the nuclear moments and reveals changes in spin dynamics.  The relaxation rate, $\lambda(T)$, obtained through exponential fits (inset of Fig.~\ref{muSR}), is very weak with a subtle increase as $T$ is decreased and is comparable to that of herbertsmithite~\cite{Mendels2007,Kermarrec2011}.

\begin{figure}	
\begin{center}
\includegraphics[width=3.35in,keepaspectratio=true]{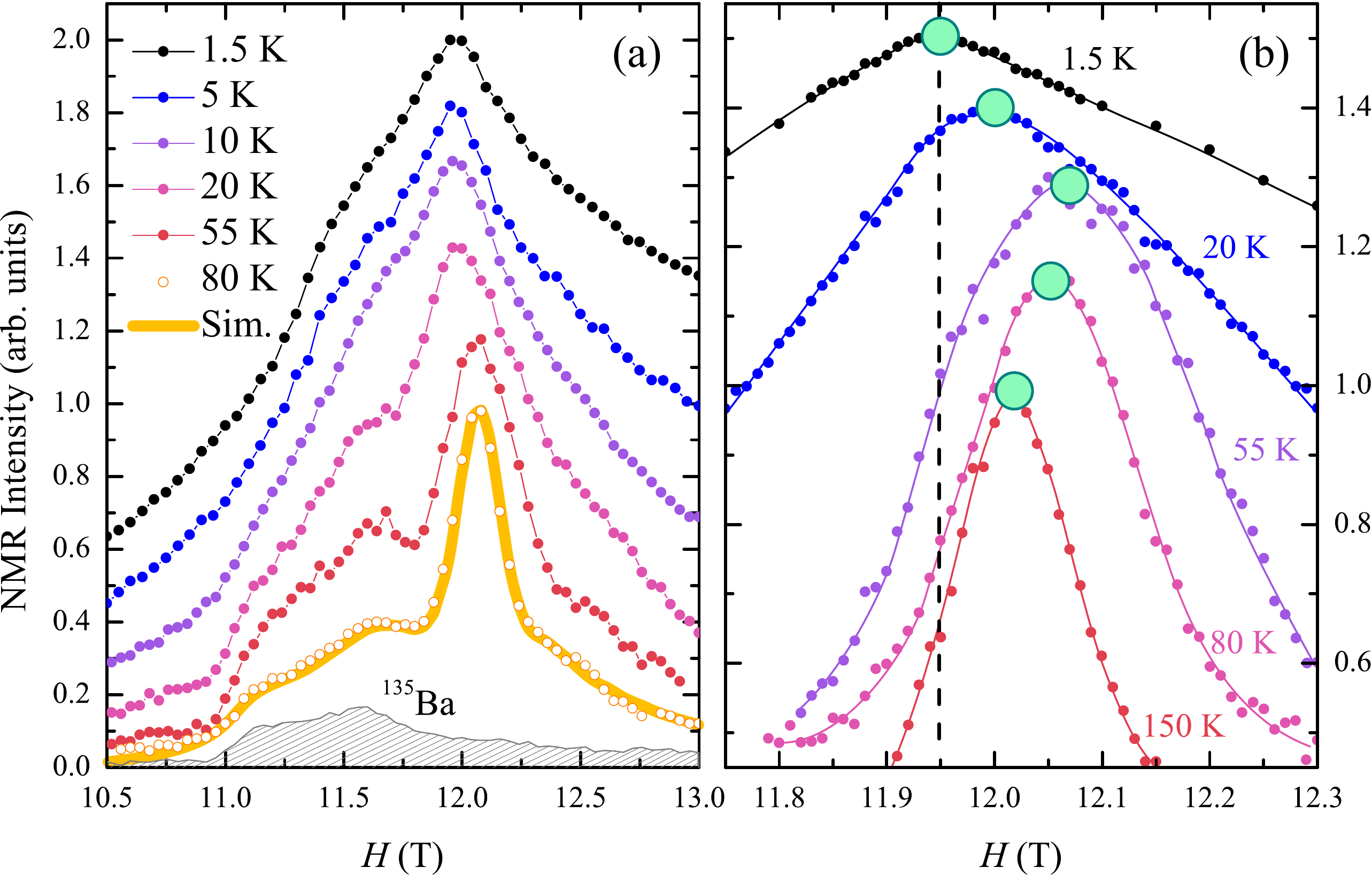}
\caption{ (a) Selected $^{123}$Sb NMR spectra, including a simulation of the spectrum at 80 K.  The $^{135}$Ba satellite can be isolated (shaded area) by roughly doubling the rf pulses. (b) Peak of spectra at several temperatures showing the non-monotonic shift (green circles).  }
\label{Spectra}
\end{center}
\end{figure}
\begin{figure}
\begin{center}
\includegraphics[width=3.35in,keepaspectratio=true]{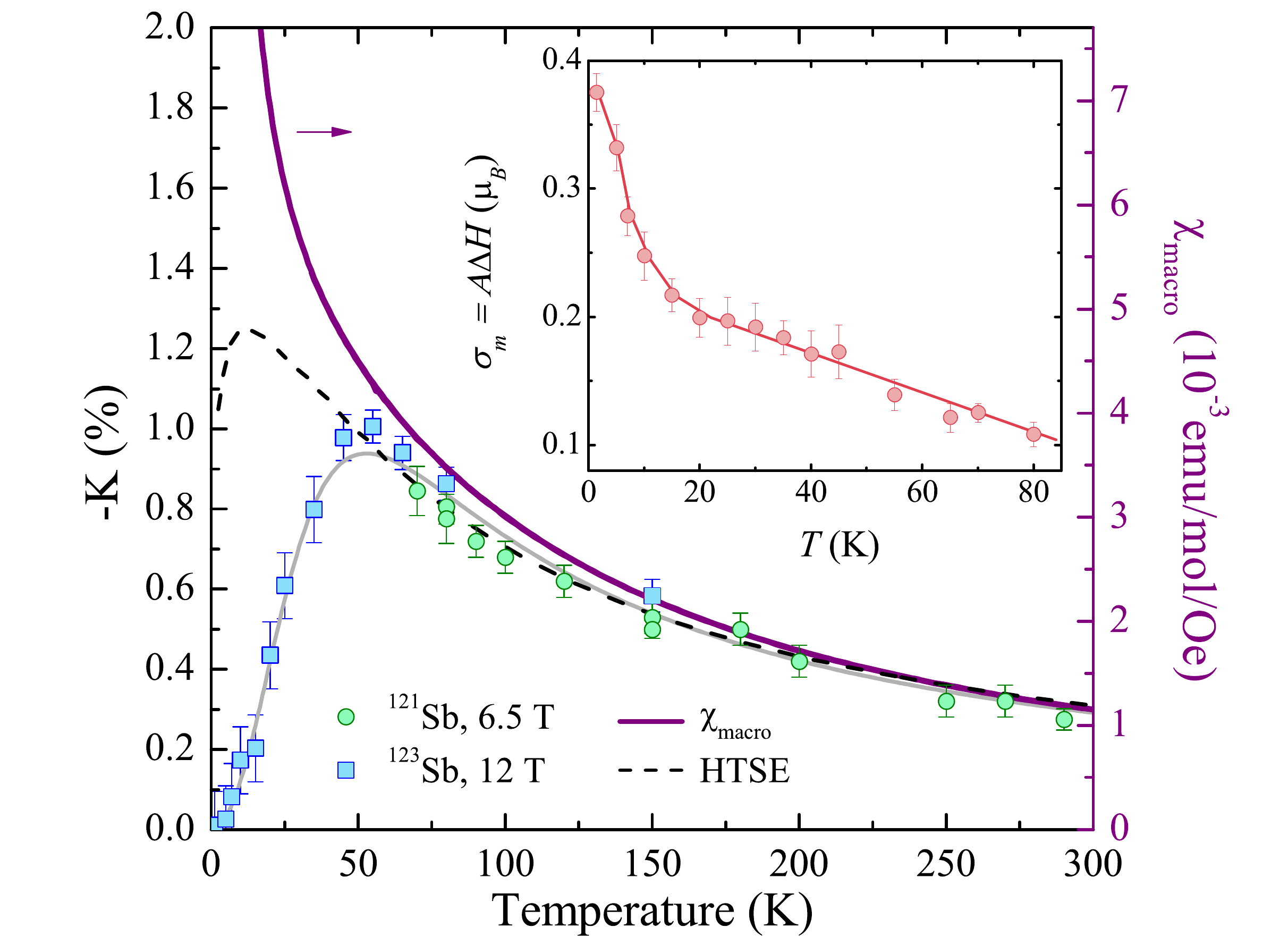}
\caption{ Left axis: NMR lineshift, $K$, taken with $^{121}$Sb at 6.4 T (high $T$ only) and $^{123}$Sb at 12 T.  Right axis: $\chi_\mathrm{macro}$ at 5 T minus diamagnetic susceptibility $\chi_\mathrm{dia}  =  -1\times 10^{-4}$ emu/mol/Oe~\cite{Nakatsuji2012}.  Also shown is the HTSE of $\chi$ for the $S=1/2$ triangular lattice antiferromagnet~\cite{Elstner1993}.  Inset:  Standard deviation of local magnetization, $\sigma_m  = A\Delta H$.}
\label{Shift}
\end{center}
\end{figure}

To precisely study the spin dynamics and in-field behavior of Ba$_3$CuSb$_2$O$_9$, we turn to a more strongly-coupled probe: Sb NMR.  Spectra are obtained with the standard $\pi/2$-$\tau$-$\pi$ pulse sequence ( $14<\tau<23$ $\mu$s) and integration of the echo as the field is swept.  Below 20 K, Cu NMR signals are seen, superimposed on the $^{121}$Sb spectra.  A $^{135}$Ba quadrupolar satellite is found near the $^{123}$Sb line, but is small and does not change in temperature. Hence, most of this work concentrates on $^{123}$Sb.  The high-temperature NMR spectra are consistent with a distribution of quadrupolar couplings, since the broadening of the quadrupolar satellites is larger than that of the central line.  This can be seen to result from Cu-Sb dumbbell disorder inherent in the decorated honeycomb lattice and random JT distortions~\cite{Nakatsuji2012}. A powder simulation with parameters $\nu_Q \simeq 3.7\pm 0.9$ MHz for the $I=7/2$ $^{123}$Sb nucleus is shown in Fig.~\ref{Spectra}.  While there are two inequivalent Sb sites in the structure, they are not discernible in the spectra. 	 The lineshift, $K$, obtained from the position of the maximum of each spectrum and is shown in Fig.~\ref{Shift}.  A scaling of $K$ at high $T$ with $\chi_\mathrm{macro}$ reveals a hyperfine coupling of $A = dK/d\chi_\mathrm{macro} \simeq12$ kOe/$\mu_B$ and shows that the temperature-independent shift (which may be chemical and/or quadrupolar) is only $\sim 80$ ppm. 


The most striking observation in the Sb NMR spectra is the non-monotonic shift (intrinsic susceptibility), $K$ ($\chi_\mathrm{int}$), shown in Fig.~\ref{Shift}.  The maximum in $\chi_\mathrm{int}$ is found at a remarkably high temperature $\sim 55$ K $\simeq \theta_W$.  Clearly this material does not behave like a nearest-neighbor $S=1/2$ triangular lattice antiferromagnet, as seen from comparison with the high-temperature series expansion (HTSE)~\cite{Elstner1993}, in Fig.~\ref{Shift}.  The appearance of strong spin correlations or dimerization at such a high temperature is unusual for a QSL and implies  a lack of magnetic frustration.

Equally important is the observation that $\chi_\mathrm{int}$ drops out appreciably and appears to approach zero as $T\rightarrow 0$.  We cannot, however, differentiate between an exponential (gapped excitations) or power-law (gapless excitations) limit with $K$ given the sizable linewidth.  $\chi_\mathrm{int}$ here contrasts heavily with behavior observed in other QSL candidates.  For example the organic triangular lattice systems, which are thought to be described by a QSL model with a Fermi surface of spinons\cite{Lee2005,Motrunich2005}, exhibit finite susceptibility as $T\rightarrow 0$~\cite{Shimizu2003,Itou2008}.

Comparison with $\chi_\mathrm{macro}$ (Fig.~\ref{Shift}) illustrates the significant level of defect contribution in the system.  Magnetization measurements at low $T$ up to 14 T, show that $\sim 16\%$ of the spins in the system are weakly interacting with $\theta_\mathrm{def} \simeq -0.5$ K at the lowest temperatures measured and are easily saturated with magnetic field~\cite{SM}.   Note that the sample studied here displays the same size of Curie tail as in Refs.~\cite{Zhou2011,Nakatsuji2012} suggesting that it might be an intrinsic property of the decorated honeycomb lattice.   Although such a weakly coupled defect or `orphan spin'~\cite{Schiffer1997} contribution is commonplace in related systems~\cite{Olariu2008,Bert2007,Okamoto2009,Quilliam2011Vesig}, here it seems likely that it originates from the out-of-plane Cu$'$ site which is inherent in the decorated honeycomb structure and is linked to the honeycomb motifs by a frustrated isosceles triangle (see the structure in Ref.~\cite{Nakatsuji2012}).
 
 Meanwhile, the linewidth (shown in the inset of Fig.~\ref{Shift}, obtained with Gaussian fits to parts of the spectra where the central line can be isolated) monotonically increases despite the drop in $\chi_\mathrm{int}$, implying that the linewidth is dominated by a defect-induced staggered magnetization or spin texture related to the Curie-tail.  At low $T$, the linewidth surprisingly represents a significant fraction of the full moment (almost 0.4 $\mu_B$), despite zero lineshift.

To assess the nature of the magnetic excitations, we turn to the spin-lattice relaxation rate, $1/T_1$.  Because the recovery curves, $M(t)$, consist of multiple stretched exponentials~\cite{Suter1998}, we have simply defined $T_1$ by the point where $1-M(t)/M_\infty = 1/e$.  The quadrupolar Sb nuclei are in principle sensitive to magnetic and charge fluctuations.  Comparison of $1/T_1$ for the two Sb isotopes (with very different gyromagnetic ratios, $\gamma$, and quadrupolar moments, $Q$) at constant $\gamma H$ and $T>20$ K, shows a ratio $^{123}T_1/^{121}T_1\simeq 1.6$.  This is close to the ratio 1.7 that we would expect if we were primarily probing magnetic fluctuations, rather than 0.6 which would be expected of charge fluctuations~\cite{Suter1998}.  Thus the JT distortions proposed in Ref.~\cite{Nakatsuji2012} are either fully static or dynamic on a time scale too fast for NMR.  

\begin{figure}
\begin{center}
\includegraphics[width=3.35in,keepaspectratio=true]{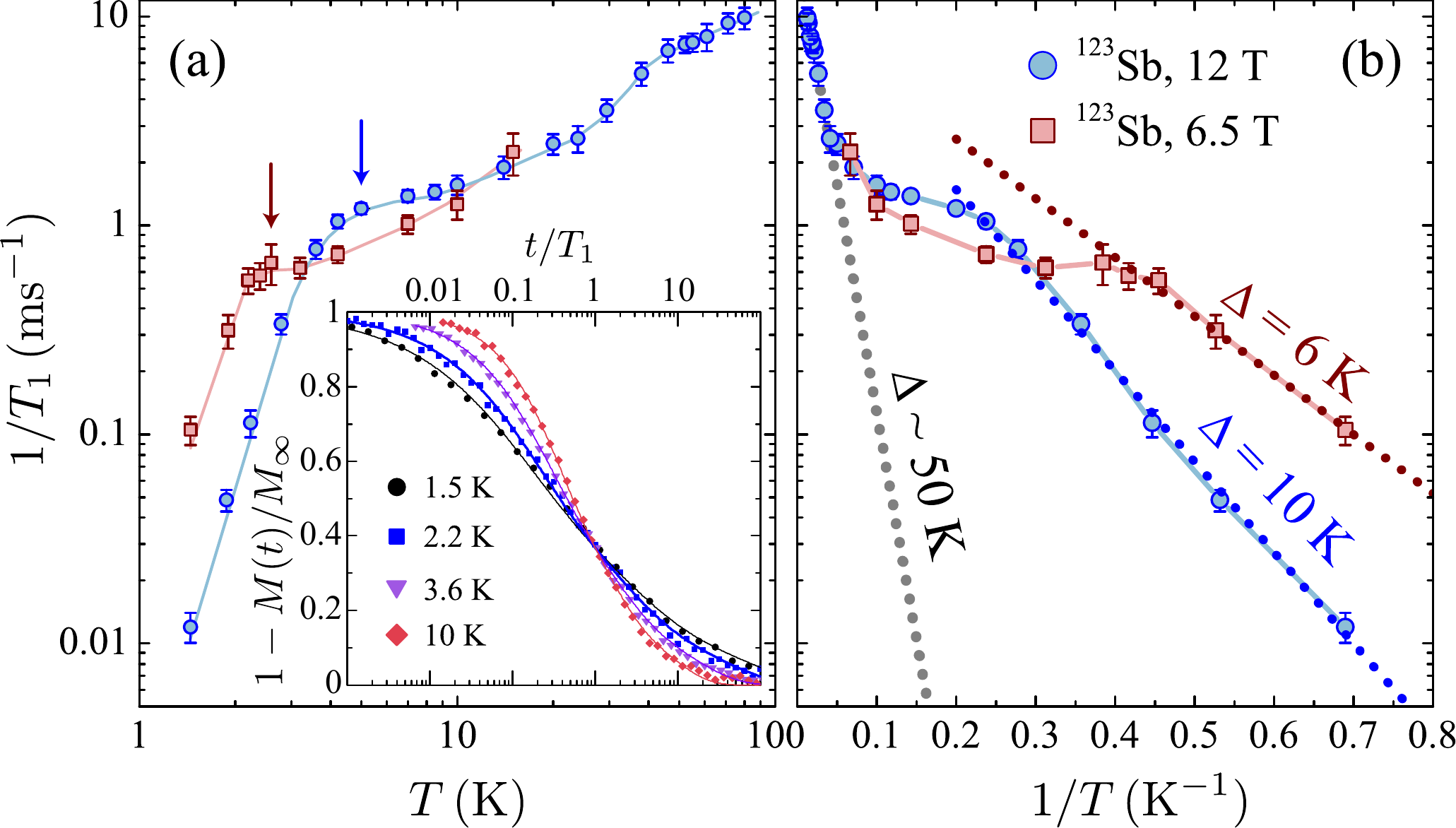}
\caption{ (a) $^{123}$Sb spin-lattice relaxation rate, $1/T_1(T)$, taken at 6.5 T and 12 T.  (b) Semilog plot of $1/T_1$ vs. $1/T$ showing gapped behavior at low temperature, with $\Delta \simeq 6$ K at 6.5 T and $\Delta \simeq 10$ K at 12 T.  Inset: the recovery curves $M(t)$ at selected temperatures showing a broadening distribution of relaxation times at low $T$, that is nonetheless small relative to the orders of magnitude drop in $1/T_1$.}
\label{T1}
\end{center}
\end{figure}

The $T$-dependence of $1/T_1$, taken at fields of 6.5 T and 12 T is shown in Fig.~\ref{T1} and shows three distinct regimes of relaxation.  At high $T$, a small drop in $1/T_1$ is observed and in a plot of $\log(1/T_1)$ vs. $1/T$ (Fig.~\ref{T1}(b)) can be linked to a $\sim 50$ K gap. This, along with the suppression of $\chi$ starting $\sim 55$ K, seems to show singlet formation, with an energy scale of $\Delta_1 \simeq J \simeq \theta_W \simeq 50$ K and is consistent with a broad peak in the inelastic neutron scattering (INS) spectra of Nakatsuji \emph{et al.}, which was attributed to the formation of short-range singlets on hexagonal motifs~\cite{Nakatsuji2012}.  However, that drop in relaxation is quickly overtaken by appreciable low energy fluctuations  giving a largely temperature independent relaxation from 5 to 20 K. This quasi-plateau in $1/T_1$ is likely related to the continuum seen in INS~\cite{Nakatsuji2012}. 

There is another, important energy scale that is driven by the applied field.  Below the quasi-plateau, a crossover to an exponential drop occurs, indicating a true gap to magnetic excitations rather than the partial dimerization at $\sim 50$ K.  The crossover temperature, $T_C$, is found to be roughly proportional to the applied field: $T_C \simeq 5$ K at 12 T and $T_C\simeq 2.6$ K at 6.5 T.  The inset of Fig.~\ref{T1}, shows the expected $T$-dependence, $1/T_1 \propto \exp(-\Delta_0/T)$, of gapped excitations and a field-dependent gap $\Delta_0(H)$.  At 12 T, $\Delta_0 \simeq 10$ K and at 6.5 T, $\Delta_0 \simeq 6$ K.  The $T_1$ recovery curves (inset of Fig.~\ref{T1}), show that below $\sim 10$ K, the distribution of relaxation times widens appreciably, suggesting that there is in fact a distribution of gaps,  as might be seen in a RSS~\cite{Shiroka2011}.

This behavior runs contrary to the case of a standard gapped dimer state or a gapped QSL where a magnetic field \emph{closes} the energy gap and eventually leads to magnetic order and power-law behavior~\cite{Giamarchi2008,Pratt2011}.  It is also surprising given the $C(T)$ of Zhou \emph{et al.}~\cite{Zhou2011} which (after subtraction Schottky anomalies) is very small but \emph{linear} in the low-$T$ limit and field-independent to as high as 9 T.  This is difficult to reconcile with our observation of a clear gap to magnetic excitations, although a theoretical model supporting such a dichotomy between $C\propto T$ and exponential $1/T_1$ has been proposed for the related $S=1$ system Ba$_3$NiSb$_2$O$_9$~\cite{Serbyn2011}.

On the basis of the structural and orbital information obtained by Nakatsuji \emph{et al.}, two main scenarios for Ba$_3$CuSb$_2$O$_9$ have been suggested~\cite{Nakatsuji2012,Balents2012}: (1) A dynamical JT distortion that allows the system to maintain isotropic character may lead to an exotic spin-orbital liquid state with gapless excitations. (2) A disordered static JT distortion may occur that stabilizes a RSS.  The configuration of distortions was proposed to obey locally a three-fold rotation axis, with three dimers around a hexagon of Cu atoms.

If the first scenario is correct, the gap seen in $1/T_1$ likely represents a transition to a frozen state with gapped magnons.  Indeed our $1/T_1$ results show qualitative similarities to the relaxation in other gapless QSL candidates~\cite{Jeong2011,Shimizu2006,Pratt2011,Itou2010} which exhibit field-induced phases with steep power law or exponential $1/T_1(T)$.  In all cases, the change in dynamics is a smooth crossover, with no peak in $1/T_1$, atypical of standard phase transitions.  A freezing of spins should be accompanied by a change in linewidth, but in this instance, it is difficult to distinguish between defect induced broadening and signs of freezing.    Much of the increase in linewidth (Fig.~\ref{Shift}) is occurring well above the crossover temperature ($\simeq 5$ K at 12 T) and is likely related to the defect contribution in $\chi_\mathrm{macro}$.

However, the QSL scenario (1) is put in doubt by strong evidence of \emph{dimerization occurring at $\sim 55$ K, as high as the exchange energy $J$}.  This is inconsistent with QSL phenomenology, where gaps are either absent or much smaller than $J$~\cite{Yan2010}, and more befitting of an unfrustrated dimer system~\cite{Sasago1997}.  Given our $T_1$ and $\chi_\mathrm{int}$ results, a more likely scenario is a type of RSS (2) with energy gap of $\Delta_1 \sim 50$ K.  As proposed by Nakatsuji \emph{et al.}~\cite{Nakatsuji2012}, this RSS consists largely of short-range singlets and our work is not consistent with a state involving arbitrarily long-range singlets~\cite{Fisher1994} that would result in a finite susceptibility at low $T$.

Clearly, however there remains a continuum of excitations below $\Delta_1$ and the field-induced gap $\Delta_0$ is much smaller.  Similarity of the field-dependent gap size, $\Delta_0/H \sim 0.8$ K/T, with the Zeeman energy for an isolated $S=1/2$ moment, $E_Z/H \simeq 1.34$ K/T, leads us to propose that it may be the weakly interacting defect spins that give rise to low-energy excitations in zero-field and that freeze out as the defects are saturated by magnetic field.  Specific heat measurements are consistent with this picture, showing Schottky anomalies that shift to higher $T$ with applied field~\cite{Zhou2011,Nakatsuji2012}.

 The effect of these defect spins on the Sb nuclei is likely indirect since no shift is seen at low $T$ and the relaxation in the plateau regime is too fast to be coming from distant spins.  With a weak coupling between defects, so $\theta_\mathrm{def} \simeq -0.5$ K, these defects would be slowly fluctuating, but even so, if they are coupled to the  nuclei with the dipolar interaction, $1/T_1$ would be an order of magnitude smaller than what is seen.  It is therefore more likely that they are distributed defects that cause a staggered magnetization or spin texture which is then strongly coupled to the Sb nuclei.  As such, the spin textures contribute heavily to the NMR linewidth, but negligibly to $K$.  This is commonplace in frustrated~\cite{Olariu2008,Quilliam2011Vesig} and gapped quantum spin systems~\cite{Alexander2010}.  The precise nature of defects and resulting spin textures remains difficult to assess but they likely result from the out-of-plane Cu$'$ site which is linked to the honeycomb motifs by a frustrated isosceles triangle (see the structure in Ref.~\cite{Nakatsuji2012}).  The Cu$'$ spins should be strongly coupled to the lattice~\cite{Nakatsuji2012}, but may become unconstrained and easily saturated under applied field as the lattice becomes dimerized at low temperatures.


To conclude, we have performed local probe measurements on the $S=1/2$ honeycomb system Ba$_3$CuSb$_2$O$_9$, revealing an exotic non-magnetic state in zero-field with dimerization occurring at 50 K and a lower temperature field-induced gap to magnetic excitations.  Our measurements primarily support a RSS  with low-energy excitations resulting from defect-induced spin textures that can be saturated with magnetic field.  This work allows us to distinguish between two proposed~\cite{Nakatsuji2012,Balents2012} orbital scenarios and likely implies random but static JT distortions.  Within this new context of short range honeycomb structural order and orbital degrees of freedom, it would be enlightening to similarly investigate the $S=1$ equivalent Ba$_3$NiSb$_2$O$_9$ which also may show a dynamic ground state~\cite{Cheng2011} but may not be JT active.

\begin{acknowledgments}
We acknowledge useful discussions with O. C\'{e}pas, S.~Nakatsuji, O.~Motrunich, P.~Deniard, T.~McQueen, A. Amato, M.~Bosio\v{c}i\'{c} and S.~Johnston.  Thanks also to D.~Dragoe for ICP analysis of our samples.  This work was partly supported by the EC FP 6 program, Contract No. RII3-CT-2003-505925.  J. Q. acknowledges the financial support of the National Science and Engineering Research Council of Canada (NSERC).
\end{acknowledgments}

\onecolumngrid
\newpage
\section{\large Supplemental Material}

\numberwithin{figure}{section}
\addtocounter{figure}{-4}

Here we provide additional information on 1) the synthesis and chemical analyses of powder samples of Ba$_3$CuSb$_2$O$_9$, 2) the analyses of powder X-ray diffraction data, 3) the details of spin-lattice relaxation and 4) high-field susceptibility measurements.
\vspace{-0.15in}
\subsection{SM.1 - Synthesis and chemical analyses of powder samples of Ba$_3$CuSb$_2$O$_9$}
Polycrystalline samples of Ba$_3$CuSb$_2$O$_9$ were prepared using a standard solid-state reaction method, starting from high-purity BaCO$_3$, CuO, and Sb$_2$O$_3$ powder materials. The dried starting materials were examined by x-ray diffraction (XRD) measurements. Stoichiometric amounts of these materials were ball-milled, pelletized, and then heated at 1100$^\circ$C for several days in air with several intermediate grindings. Samples were furnace cooled at the end of the final heat treatment. As-prepared powders are yellow-brown in color. The samples were analyzed with a scanning electron microscope (SEM) equipped with an energy-dispersive spectroscopy (EDS) detector. SEM analyses done in both backscattered and secondary electron imaging modes showed the samples to be homogeneous. Semiquantitative microprobe elemental EDS analysis was performed at different positions on the sample surfaces using polished sections. Atomic ratios of cations were found to be 1.48(3), 2.9(2), and 0.51(3) for Ba/Sb, Ba/Cu, and Cu/Sb ratios, respectively. Cation stoichiometries were also determined by means of inductively coupled plasma optical emission spectroscopy (ICP-OES) which yielded atomic ratios of 1.44(1), 3.02(3), and 0.48(1) for Ba/Sb, Ba/Cu, and Cu/Sb ratios, respectively.  These ratios are close to the nominal ones. As the deviation from 2:1 Sb/Cu stoichiometry is well below 10\%, the results of Nakatsuji \emph{et al.} [3] suggest that most of the sample should remain in its hexagonal form (not undergoing an orthorhombic Jahn-Teller distortion) down to low temperatures. A thermogravimetric study was performed in air and in oxygen atmosphere, and no weight decrease or increase was observed below 1200$^\circ$C.	

\begin{figure}[b]
\begin{center}
\includegraphics[height=3.5in,keepaspectratio=true]{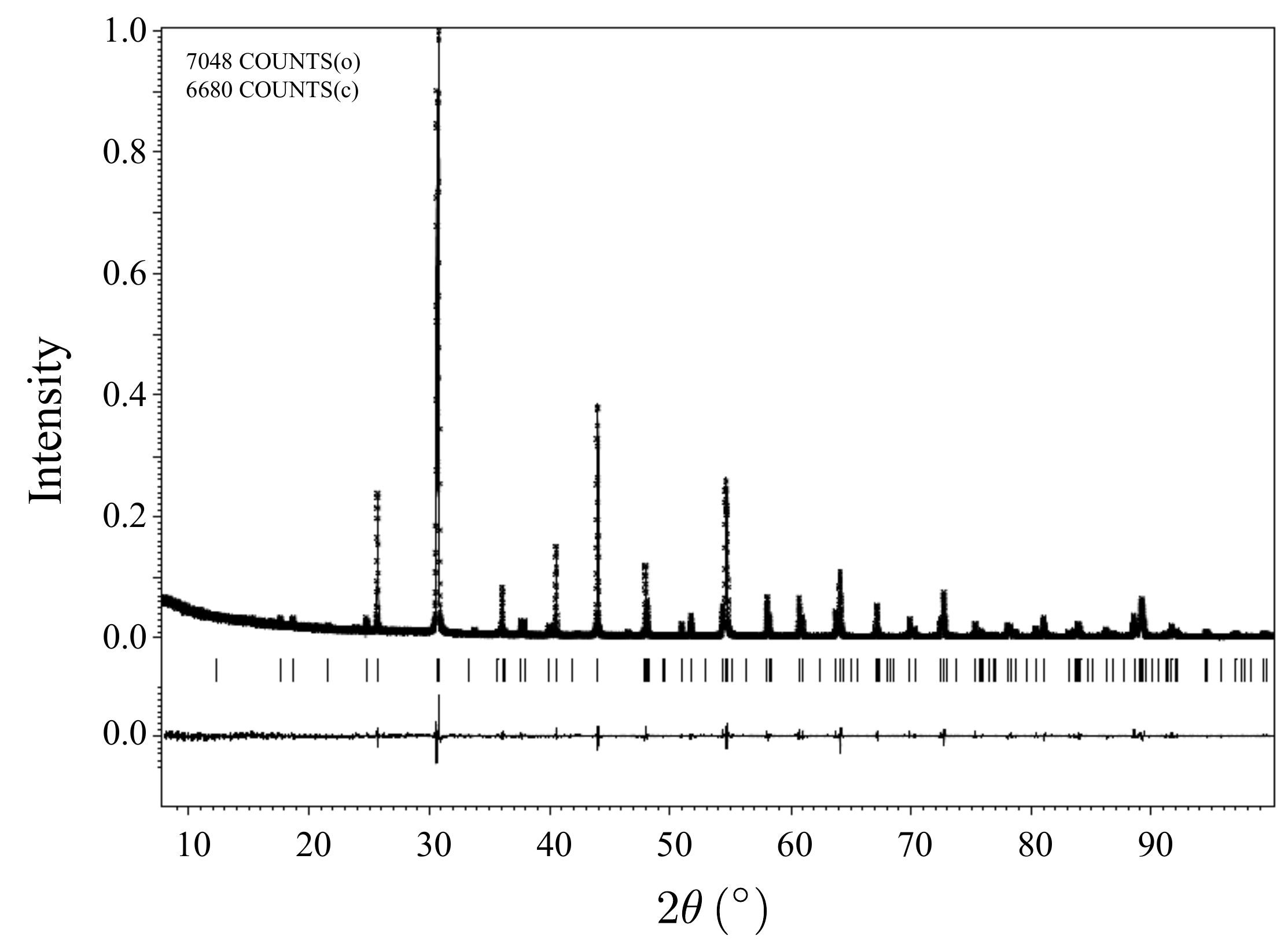}
\caption{Observed and calculated (solid curve) powder XRD pattern of a sample of Ba$_3$CuSb$_2$O$_9$ collected at room temperature using Cu K$\alpha_1$ radiation. The tickmarks indicate the positions of the Bragg reflections. The lower curve shows the difference between the observed and calculated data on the same scale. 
\label{SIXRD}
}
\end{center}
\end{figure}

\subsection*{SM.2 -- Powder X-ray diffraction data for Ba$_3$CuSb$_2$O$_9$}
Laboratory powder XRD data were collected at room temperature in the $2\theta$ range of 8-100$^\circ$ using Cu K$\alpha_1$ radiation ($\lambda=1.5406$ \AA). All patterns showed narrow diffraction peaks and no sign of unreacted starting materials (Note that the most intense peaks of Tenorite do not coincide with those of the desired phase). As noticed earlier for other Ba$_3M$Sb$_2$O$_9$ ($M$ = Mn, Co, Ni) compounds~\cite{Doi}, a very small amount of impurity phase BaSb$_2$O$_6$ (JCDS \#82-0520) showed up in the diffraction patterns. The data were analyzed with the Rietveld method by use of the JANA program~\cite{JANA}. Three $2\theta$ regions where the BaSb$_2$O$_6$ peaks showed up were excluded from refinements. We used the recently published $P6_3/mmc$ hexagonal crystal structure~\cite{Nakatsuji} as a starting model. The lattice parameters, $a =$ 5.80390(4) $\mathrm{\AA}$ and $c =$ 14.3231(1) $\mathrm{\AA}$ were in good agreement with those determined from the single-crystal study in Ref.~\cite{Nakatsuji}. There was no sign of additional structural periodicities. For the Rietveld analysis, independent anisotropic atomic displacement parameters (ADPs) for each Ba site, and isotropic ADPs for the other sites, were refined with full occupancies of all sites. The two central sites of the face sharing octahedra (Wykoff $4f$ positions) were equally occupied by Cu and Sb atoms. With this model, patterns were refined with reliability factors of $R_p = 0.0855$, $R_{wp} = 0.1225$ and GOF = 1.2. Maximum and minimum electron densities in the difference Fourier map were 0.8 electron/\AA$^3$ and -1.2 electron/\AA$^3$, respectively. Empirical bond-valence sums for Ba1, Ba2, Sb1, O1, and O2 atoms were consistent with nominal oxidation states Ba$^{2+}$, Sb$^{5+}$, O$^{2-}$. We were not able to resolve aÊfew percent change in Cu/Sb2 or Cu/Sb1 ratios nor a weak oxygen off-stoichiometry. A representative Rietveld profile fit is shown in Fig. SM1.

\subsection*{SM.3 -- Isotopic dependence of spin-lattice relaxation}

\begin{figure}[b]
\begin{center}
\includegraphics[height=2.5in,keepaspectratio=true]{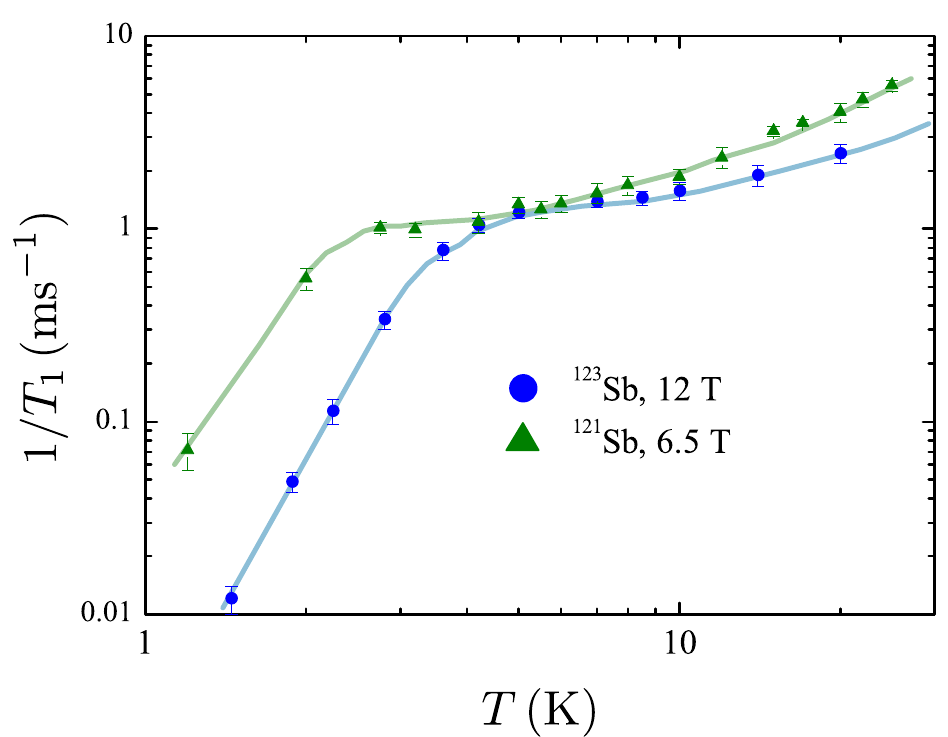}
\caption{$T_1$ vs. temperature for the two different isotopes of Sb taken at the same frequency, $\gamma H = 66$ MHz.
\label{SIT1}
}
\end{center}
\end{figure}

In Fig. SM2, $1/T_1(T)$ is shown for the two different Sb isotopes at constant frequency $\gamma H =  66$ MHz.  This comparison allows us to roughly determine whether our relaxation is dominated by magnetic fluctuations or by charge fluctuations, both of which can affect quadrupolar nuclei.  The multi-exponent relaxation of a quadrupolar nucleus yields different results depending on the spin number $I$.  Here $I=5/2$ for the $^{121}$Sb isotope and $I=7/2$ for the $^{123}$Sb isotope.  The parameters of Ref.~\cite{Suter} illustrate that $^{121}T_{1e}/^{123}T_{1e} \simeq 2W_{121}/W_{123}$ where $W_n$ is the spectral density of fluctuating fields for isotope $n$.  Following the calculations of Moriya~\cite{Moriya}, $W_n \propto \gamma_n^2$.  Here $\gamma_{121} / \gamma_{123} = 1.85$.  Thus if fluctuations are magnetic in origin, we expect roughly a factor of 1.7 decrease in $1/T_{1e}$ as we move from $^{121}$Sb to $^{123}$Sb.  Indeed, this is very close to what is observed at higher temperatures (away from any field induced physics), as can be seen from Fig. SM2(a).  

On the other hand, in the case of purely quadrupolar relaxation, we should expect the isotopic dependence to vary in the opposite manner as $^{121}W_Q/^{123}W_Q \simeq (^{123}Q/^{121}Q )^2 = 1.65$.  This is clearly not the case thus we are primarily sensitive to magnetic fluctuations and not to charge fluctuations.  Either the charge fluctuations are completely static (frozen) in the temperature range of our measurements, or else in the case of a dynamic Jahn-Teller effect~\cite{Nakatsuji}, the charge fluctuations may occur at too high a frequency to be probed with NMR.

\subsection*{SM.4 - High-field magnetization measurements}

High-field magnetization measurements were performed with a Cryogenic vibrating sample magnetometer (VSM) system, at temperatures from 2 K up to 5 K.  The resulting curves, shown in Fig. SM3, are fit with the formula
\begin{equation}
M(H) = n_\mathrm{imp} (g/2)\mu_B \tanh\left[ \frac{g\mu_B H}{k_B(T + \theta_\mathrm{imp})}\right] + \chi_\mathrm{res}H 
\end{equation}
where $\theta_\mathrm{imp}$, and $\chi_\mathrm{res}$ are left as free parameters and $g$ is taken from ESR measurements of Ref.~\cite{Nakatsuji}.  The total number of Cu spins in the system, $n_\mathrm{tot}$, is taken from a high temperature Curie-Weiss fit using the same $g$-factor.  The result is that $n_\mathrm {imp} / n_\mathrm{tot} = 0.16$, showing that 16\% of the spins in the material are easily saturated, with $\theta_\mathrm{imp}$ varying from -2.4 K (at $T=5$ K) to -0.5 K (at $T=2$ K).  There remains some residual susceptibility, $\chi_\mathrm{res}$ that is mostly constant in $T$ and averages to $\sim 2.8\times 10^{-3}$ emu/mol/Oe coming from spins that are not easily saturated.

Furthermore, we note that there are no features in the magnetization indicative of a phase transition occurring at the same field/temperature as the field-induced gap seen in NMR. 

\begin{figure}
\begin{center}
\includegraphics[height=2.5in,keepaspectratio=true]{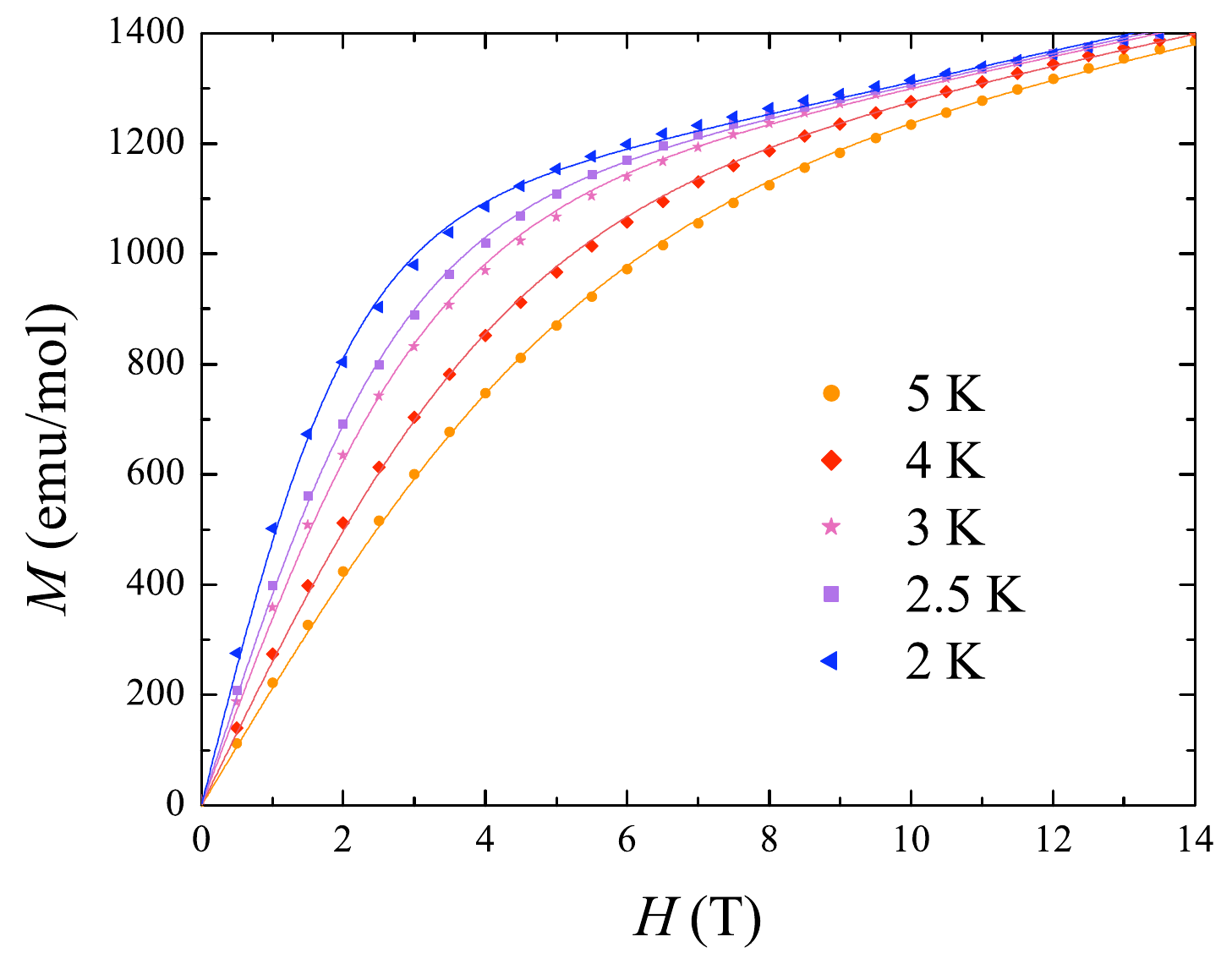}
\caption{High field magnetization measurements of Ba$_3$CuSb$_2$O$_9$.  Fits are described in the text.
\label{VSM}
}
\end{center}
\end{figure}

\vspace{0.2in}
We thank D.~Dragoe for ICP analysis of our samples and P. Deniard for useful discussions regarding the Rietveld analysis.


\begin{thebibliography}{10}
\bibitem{parimp}
A small, $\sim 4\%$, quickly relaxing contribution to the $\mu$SR asymmetry is seen at all temperatures and has been subtracted. This is likely a small ferromagnetic impurity phase which is also seen in low-field susceptibility measurements.
\bibitem{upperbound}
A conservative upper bound on the size of possible frozen moment  of $8\times 10^{-4}\mu_B$ may be obtained by assuming the change in muon depolarization between 10 K and 20 mK is a result of static magnetism.
\end{thebibliography}
\end{document}